\documentstyle[12pt,psfig]{article}

\begin{document}
\begin{center}
{\large {\bf{PION DISSOCIATION IN HOT QUARK MEDIUM}}}
\vskip 36pt
{\bf Abhijit Bhattacharyya\footnote{Electronic Mail : 
phys@boseinst.ernet.in}, Sanjay K. Ghosh\footnote{Electronic Mail : 
phys@boseinst.ernet.in} and Sibaji Raha\footnote{Electronic 
Mail : sibaji@boseinst.ernet.in}} 
\vskip 1pt
Department of Physics, Bose Institute\\
93/1, A.P.C.Road\\
Calcutta 700 009, India
\end{center}
\vskip 36pt
\begin{abstract}
Pion dissociation in a medium of hot quark matter is studied.
The decay width of pion is found to be large but finite at temperatures much higher than the so called critical temperature of 
chiral or deconfinement transition. Consequently, pions
should coexist with quarks and gluons at such high temperatures.
The result is in agreement with the lattice calculations. The implication of 
the above result in the study of Quark-Gluon plasma is discussed. 
\end{abstract}
\vskip 25pt
\indent
A strong prediction of Quantum Chromodynamics (QCD), {\it the underlying 
theory of
strong interaction}, is that at very high temperature and/or density, the
bulk properties of strongly interacting matter would be governed by the
quarks and gluons, rather than the usual hadrons. Such a phase is called
quark gluon plasma (QGP) \cite{a} in the literature and the search for 
such a novel phase of matter constitutes a major area of current research 
in the field of high energy physics.
\par
The properties and dynamics of QGP are obviously governed by QCD. This 
conceptually 
straight forward task is, however, quite formidable in practice, particularly
because of the failure of perturbative QCD already in the temperature
range in the vicinity of $\Lambda_{QCD}$ ($\sim$ few hundred MeV) \cite{a1}. 
Analytical non-perturbative methods are not yet sufficiently developed
to be of much use in this context and as such, the lattice  formulation of
QCD has developed into the primary vehicle for the study of
QGP \cite{c}. In addition to the intensive 
computation, both in
terms of CPU time and numerical complexities, one can only address static
properties in the lattice. As a result, the space - time evolution of
the system formed in the ultrarelativistic heavy ion collisions remains
unapproachable in the framework of the lattice; thus the alternate, classical
picture of hydrodynamic evolution, which accounts for the overall energy -
momentum conservation in a collective manner and not much else, has been
used quite extensively to study the evolution of the QGP \cite{d}.
QCD inputs enter into such a picture through the equation of state 
of the QGP, preferably evaluated on the lattice ( but more often, through
a phenomenological bag model \cite{b}). 
\par
An inescapable feature of the collision
process is that the quarks and gluons must, at some epoch, turn into 
hadrons which would ultimately be detected, never the individual 
quarks and gluons. 
The actual process of hadronisation, however, continues to elude us. It has 
been 
widely postulated that there could be an actual phase transition (the order 
of which is an open issue), separating the QGP phase from the hadronic
phase \cite{f}. The recent results, showing the lack of thermodynamic 
equilibrium \cite{g} in the quark-gluon phase in ultrareletivistic heavy
ion collisions, indicate that such an ideal situation is unlikely. It
should also be noted at this juncture that although the persistence of
non-perturbative effects till very high temperatures was suggested in the 
literature quite early on \cite{h}, it is only recently that the lattice
results have confirmed that non-perturbative hadron like excitations 
could survive at temperatures far above the chiral phase transition 
temperature \cite{i}. It is thus imperative to understand the behaviour of
such hadronic resonances, their formation, stability and so on, in a
quark gluon medium at high temperature. In this work we confine
our attention to the case of pions alone, which is naturally a 
prototype of all hadrons. Being the lightest hadron, pions account for 
the bulk of the multiplicity. 
\par
Formation of pions, a bound state of light relativistic quarks, is an
extremely difficult problem to handle in QCD. This is where all the
troublesome features of non-perturbative QCD would make their presence
felt. We therefore employ the usual 
practice of looking at the pion as a Goldstone boson arising from the 
spontaneous breaking of the chiral symmetry. The coupling of the pion to the 
quarks can then be obtained in a straightforward manner, by starting 
with the free lagrangian, imposing a chiral transformation and demanding 
invariance under such a transformation. 
Explicitly, the chiral angle is associated with 
the pion field and the quark field is rotated by the chiral \cite{k} 
transformation,
\begin{equation}
q^{'}= exp \left[i {{\vec\pi.{\vec \tau} \gamma^{5}}\over {2 f_{\pi}}}\right ] q
\label{eq:1}
\end{equation}
Expanding the exponential to first power in ${1 \over {f_\pi}}$, we obtain 
the pion-quark coupling. The interaction term is given by \cite{k},
\begin{equation}
{\cal L}_{int} =  {{m_q} \over {f_\pi}}{\bar q} \gamma_5 {\vec \tau} .
{\vec \pi} q
\label{eq:intt}
\end{equation}
where $m_q$ is the quark mass, $f_{\pi}$ is the usual pion decay
constant (=93 MeV), $q$ is quark field and $\tau$ is the usual 
Pauli matrix.
\vskip 0.3in
\begin{tabular}{cl}
\psfig{file=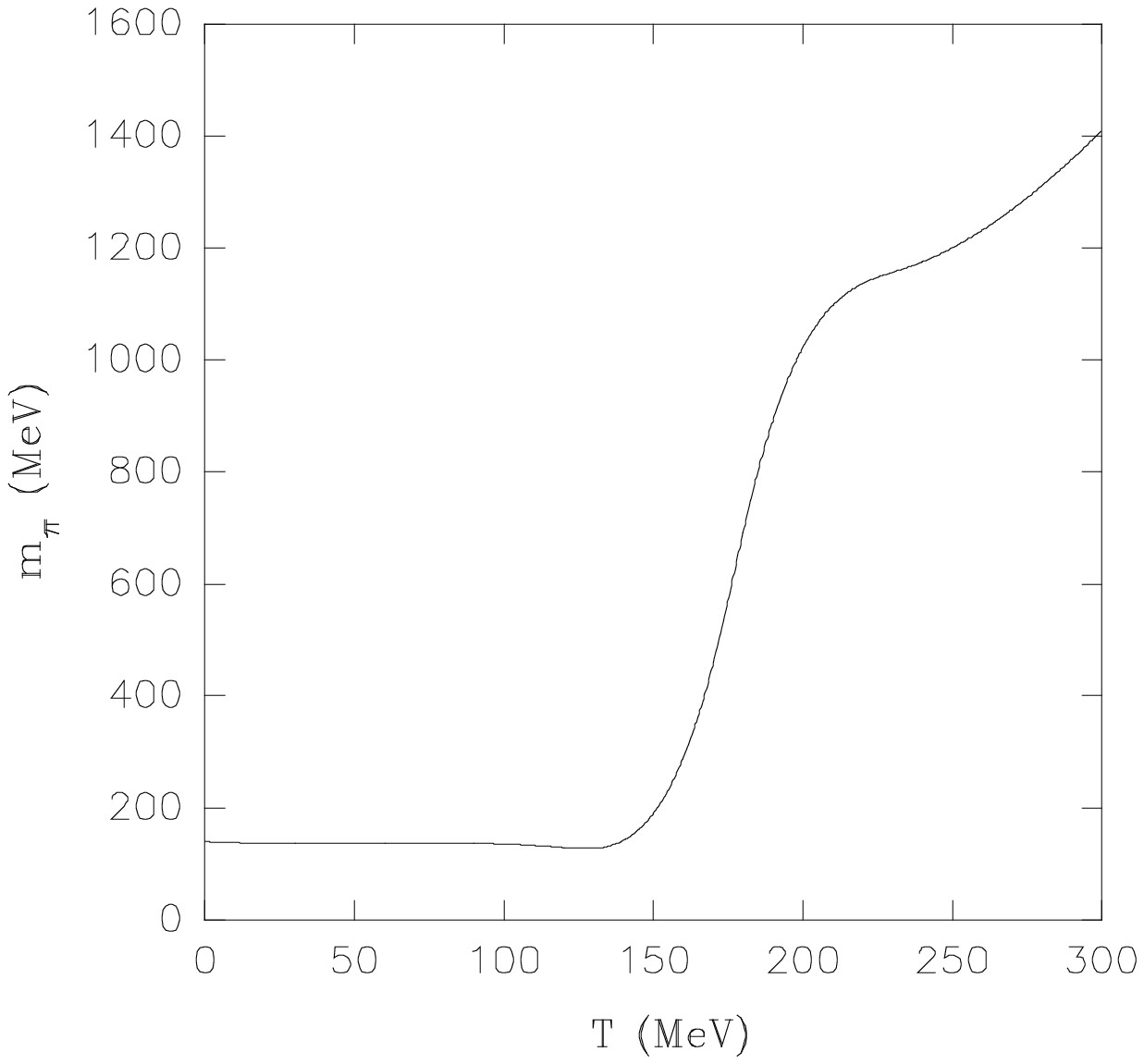,width=5in,height=5in}
\end{tabular}
\vskip 0.2in
\centerline {{ Figure 1 : Temperature dependence of pion mass from lattice 
data.}}
\vskip 0.3in
\par
Even with such an interaction, the formation of pions from quarks and
gluons would require an involved analysis through the Bethe-Salpeter
equation. Such a study is very much on our agenda but we do not address
this issue here. In the present work, we concern ourselves with the decay
of pionic excitations, the properties of which we assume to be given by
the lattice calculations. It should be reiterated that at
temperatures above the critical temperature, these pionic excitations are
more like resonances with large effective masses \cite{i,j}. The variation of
the pion mass with temperature, as calculated in the lattice \cite{j}, 
is shown in figure 1. In the following, we study the decay width of such
pionic excitations in the hot quark medium as a function of temperature,
starting with the interaction given above in equation (\ref{eq:intt}). 
\par
The quark mass $m_{q}$ appearing in eq. (\ref{eq:intt}) is a very important
ingredient in our calculation. In the absence of any medium and/or
dynamic effect, $m_{q}$ should assume the value of the current quark mass. 
On the other hand, we know that due to the spontaneous breakdown of the
chiral symmetry, quarks attain the value of the constituent quark mass \cite{k}.
However, we are investigating the behaviour
of pions in quark medium at very high temperatures ($>>$ chiral symmetry
restoration temperature)
where quarks pick up a large thermal mass \cite{l} due to medium effects. 
So, in our calculation, we have taken, for the sake of completeness,
three different values for $m_{q}$, 
namely the current quark mass ($\sim 10$MeV), the constituent quark mass 
($\sim 300$MeV) 
as well as the thermal mass (see below).
\par
The decay width of a pion in its rest frame is given by the usual expression, 
\begin{equation}
\Gamma = \int {{d^3p_1} \over {2p_1^0(2\pi)^3}} 
{{d^3p_2} \over {2p_2^0(2\pi)^3}} {{(2\pi)^4 \delta^4(Q - p_1 - p_2)} 
\over {2Q_0}} |M|^2 (1-f(p_{1})) (1-f(p_{2}))
\end{equation}
where $M$ is the matrix element, $Q$ is the momentum of the pion and $p_1$ and 
$p_2$ are the momenta of $q$ and ${\bar q}$. $f(p_1)$ and $f(p_2)$ are
the usual Fermi-Dirac distribution functions accounting for the
Pauli blocking of final state quarks. The matrix element is given 
by
\begin{equation}
M={{m_q} \over {f_\pi}}\bar {q} \gamma_{5} q \tau
\end{equation}
From (3) and (4), the final expression for the 
decay width of the pion to a quark anti-quark pair, in the rest frame of the 
pion, is given by,
\begin{equation}
\Gamma = {1 \over {4\pi {m_{\pi}}^2}} \left[{m_{q}\over f_{\pi}}\right]^{2} 
({m_{\pi}}^{2} - 4 {m_{q}}^{2})^{3/2}
(1 - f(E_{1})) (1 - f(E_{2}))
\end{equation}
where $E_{1}$ and $E_{2}$ are the quark energies. The $\delta$-function 
in equation (3) yields $E_1 = E_2 = m_{\pi}/2$.
\par
The thermal quark mass is defined as \cite{l}
\begin{equation}
m_{th} = \sqrt{m_{curr}^2 + {{g_s^2T^2} \over 9}}
\end{equation}
where \cite{l}
\begin{eqnarray}
g_s &=& \sqrt{4 \pi \alpha_s} \nonumber\\
\alpha_s &=& {{6 \pi} \over 29} ln \left({{3T} \over \Lambda}\right)
\end{eqnarray}
In the above equation, $\Lambda$ is the QCD parameter and $m_{curr}$ 
the current quark mass. We have considered three values of 
$\Lambda$, $0.3$ , $0.2$ and $0.1$ GeV.
The variation of the pion decay width with temperature for different quark 
masses
and $\Lambda$ are shown in figure 2. The decay width for the current quark mass
is not included in the figure, as for this case, the decay width is 
very small; in fact it is practically zero in the scale of the present 
figure.
\par
Figure 2 shows that the decay width is very high at high temperature 
($\sim 0.3$ GeV) and decreases with decreasing temperature, going to zero 
at around $T = 0.16$ GeV. It is worth noticing that at around the same 
temperature, the effective pion mass attains the value of the free pion 
mass (figure 1). The dependence on $\Lambda$ is very clear. 
The decay width increases with increase in $\Lambda$. However, the 
temperature at which the decay width goes to zero does not depend sensitively
on $\Lambda$. 
The decay width is found to be maximum for the constituent quark mass.
The fact that for the current quark mass the decay width is very small,  
whereas for the constituent quark mass it is fairly high, leads one  
to infer that the decay width decreases as the quark mass decreases. 
This explains the dependence on $\Lambda$ too, as larger $\Lambda$ 
corresponds to larger $m_q$ at any given temperature. In figure 2 one 
should also note the 'shoulder' like structure in decay width
around $T = 200 MeV$. This 'shoulder' is a reflection of the temperature 
variation of pion mass (figure 1), which also has a 'shoulder' around 
that temperature.     
\vskip 0.3in
\begin{tabular}{cl}
\psfig{file=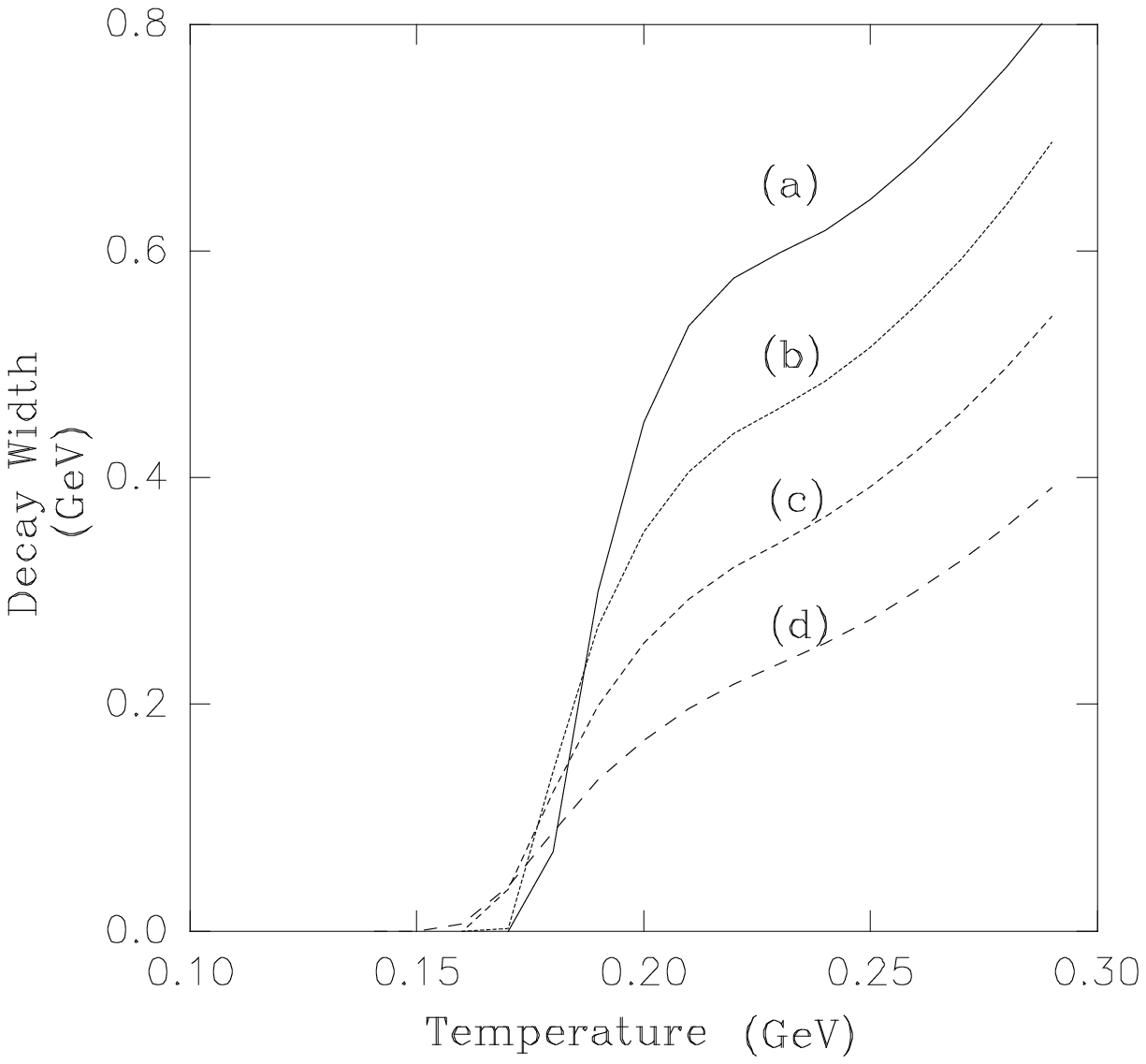,width=5in,height=5in}
\end{tabular}
\vskip 0.2in
\centerline { Figure 2 : Temperature dependence of pion decay width; 
(a) for the}
\centerline{constituent quark mass, (b), (c) and (d) are for the thermal}
\centerline{ quark masses for $\Lambda = 0.3,$ 
 $0.2$ and $0.1$ GeV respectively.}
\vskip 0.3in
\par
The increases of the pion decay width with increasing $m_q$ is somewhat 
against the common notion that as the difference in the 
masses of pion and quark increases the decay width should increase. For the  
interaction used here, the decay width is proportional to $m_q^2$ 
(equation 2) and as a result the decay 
width increases with the increase in quark mass. The decay width is 
maximum when the quark mass is taken to be its constituent mass {\it {i.e.}}
Its value at $T = 290$ MeV is $0.8$ GeV; as a result, the pions formed at that 
temperature will decay immediately to $q {\bar q}$ pairs. 
\par
Our results will have a strong bearing on the study of hadronisation. As already 
mentioned, the lack of thermodynamic equilibrium in a QGP system implies that 
one may not get a clear-cut phase transition from QGP to hadrons. Thus,  
to understand the process of hadronisation, one should really start from 
a very high temperature ($>>$ expected $T_c$) and then let the system evolve 
dynamically towards  lower temperatures. Here what one would find, as 
indicated from our present calculation, is that initially a very small number of 
pions would be present in the system along with quarks and gluons.  
Then, even if additional pions are formed through $q {\bar q}$ fusion 
and/or bound state formation, the total number of pions should not increase 
very fast, as most of them must decay immediately due to the large 
decay width at such high temperatures. Only in the 
vicinity of $T \sim 170 MeV$, where the decay width is small, the number 
of pions would start increasing significantly and gradually become dominant 
compared to the number of quarks at some lower temperature. However, the  
exact value of the temperature, at which the decay width goes to zero, 
 will depend on the value of the quark mass considered.
\par
To summarise, we have calculated, for the first time, the decay of pions,
which is a prototype of all hadrons, in a hot quark medium. The most
interesting and noteworthy feature is that, even without any consideration
of the detailed evolution and dynamics of the system, the pionic modes are
found to dominate around a temperature of $160$ MeV. Though the question 
whether 
this is a signature of a phase transition cannot be addressed within the
framework of the present work, the fact that most of the pions
decay into quarks, owing to a large decay width at temperatures higher than 
$T_c$, is a remarkable finding.
\par
It will be interesting to compare the pion decay width obtained here 
with the decay widths of other mesons. Qualitatively, the same conclusion
should hold but it remains to be seen if all hadronic modes start becoming
important at about the same temperature. 
Work in this direction is in progress.  
\par
The work of AB and SKG have been supported, in part, by the Department of 
Atomic Energy (Government of India) and Council of Scientific and Industrial 
Research (Government of India), respectively.  


\vskip 0.4in
\begin{thebibliography}{99} 
\bibitem{a} E.V.Shuryak, {\it Phys. Rep.} {\bf 61}, 71 (1980).
\bibitem{a1} See, for example, {\it Quark-Gluon Plasma}, Ed. R.C.Hwa, World 
Scientific, Singapore (1990)
\bibitem{c} F.Karsch, {\it Z. Phys.} {\bf C38}, 147 (1988).
\bibitem{d} J.Alam, S.Raha and B.Sinha, {\it Phys. Rep.} {\bf 273},    
243 (1996). 
\bibitem{b} P.Hasenfratz and J.Kuti, {\it Phys. Rep.} {\bf 40}, 73 (1978).
\bibitem{f} It should be noted that, in the context of QGP or hot hadronic
matter, one comes across two kinds of phase transitions, namely, the chiral 
phase transition and the deconfinement phase transition. The current consensus 
is that although these two phase transitions are certainly distinct from 
each other, the critical temperatures may not be much different. We have taken 
them to be same for the present purpose.
\bibitem{g} E.V.Shuryak, {\it Phys. Rev. Lett.} {\bf 68}, 3270 (1992); 
S.Chakraborty, S.Raha and B.Sinha, {\it Mod. Phys. Lett.} {\bf A7}, 927 
(1992); J.Alam, S.Raha and B.Sinha, {\it Phys. Rev. Lett.} {\bf 73}, 1895 
(1994); T.S.Biro, B.M{\"u}ller, M.H.Thoma and X.N.Wang, {\it Nucl. Phys.} 
{\bf A566}, 543c (1994).  
\bibitem{h} M.Pl{\"u}mer, S.Raha and R.M.Weiner, {\it Phys. Lett.}  {\bf B139}, 
198 (1984); {\it Nucl. Phys.} {\bf A418}, 549c (1984).  
\bibitem{i} K.Born {\it et al.}, {\it Phys. Rev. Lett} {\bf 67}, 302 (1991). 
\bibitem{k} T.Frederico and G.A.Miller, {\it Phys. Rev.} {\bf D45}, 4207 
(1992).
\bibitem{j} A.Gocksh, {\it Phys.Rev.Lett.} {\bf 67}, 1701 (1991).
\bibitem{l} T.Altherr and D.Seibert, {\it Phys. Rev.} {\bf C49}, 1684 (1994).  
\end{thebibliography}
\end{document}